\documentclass[aps,prb,twocolumn,groupedaddress,floatfix]{revtex4-1}
\usepackage{graphicx}% Include figure files
\usepackage{dcolumn}% Align table columns on decimal point
\usepackage{bm}% bold math
\usepackage{amsmath}
\usepackage{color}
\begin{document}
%chiral effect in vesalago lens focusing: armchair vs zigzag graphene p-n junction
\title{Band engineering and elastic properties of strained armchair graphene nanoribbons: semiconductor vs metallic characteristics
}
\author{Sanjay Prabhakar$^1$ and Roderick Melnik$^{1,2}$ }
\affiliation{
$^1$The MS2Discovery Interdisciplinary Research Institute, M2NeT Laboratory, $^2$Wilfrid Laurier University, Waterloo, ON N2L 3C5, Canada\\
BCAM, Alameda Mazarredo 14, 48080 Bilbao, Spain
}
\date{January 02, 2019}

\begin{abstract}
An odd number of zigzag edges in armchair graphene nanoribbons and their mechanical properties (e.g., Young's modulus, Poisson ratio and shear modulus) have potential interest for bandgap engineering in graphene based optoelectronic devices. In this paper, we consider armchair graphene nanoribbons passivated with hydrogen  at the armchair edges and then apply the strain for tuning the bandgaps. Using density functional theory calculations, our study finds that the precise control of strain can allow tuning the bandgap from semiconductor to mettalic and then again switching  back to semiconductor.  In addition, we also show that the strained graphene nanoribbon passivated with hydrogen molecules can have  large out-of-plane deformations demonstrating the properties of relaxed shape graphene. We express the strain induced by hydrogen in terms of binding energy. Finally, we characterise the effect of strain on the mechanical properties that can be used for making straintronic devices based on graphene nanoribbons.
\end{abstract}

%\pacs{71.70.Ej, 73.61.Ey, 85.75.Hh}
%\keywords{}


\maketitle

\section{Introduction}

The thinnest two dimensional (2D) materials, e.g.,  graphene, MoSe$_2$ and other TMDs materials have potential interest for making next generation optoelectric, spintronic, straintronic devices that might have performance better than conventional silicon devices~\cite{androulidakis18,mannix18,wu18,feng18,hapuarachchi18,azar18,geim-nature07,coleman-science11,savage-nature12,manzeli17-TMD}. For example, the measurement of charge mobility in the graphene CMOS devices is far better than the best silicon
devices~\cite{savage-nature12,novoselov04,liao-nature10,yang17,friedman17,altintacs17}. Evidence from experimental measurements confirmed that  2D materials possess unique physical properties, e.g. half integer quantum Hall effect, non-zero Berry curvature and Zak's phase, high mobility charge carriers~\cite{novoselov05a,novoselov05,novoselov04,savage-nature12,cao17,nguyen17}.
Although graphene as a material is exciting at the level of making optoelectronic devices, its bandgap opening is still one of the biggest challenge, e.g. graphene does not have any bandgaps at the Dirac point.
However, by using several state-of-the-art techniques and possibility of considering different 2D materials, e.g. MoSe$_2$, one can easily control the bandgaps. Also, a  small bandgap opening in graphene  is achieved by considering the effect of spin-orbit coupling, strain and magnetic field. One can also make optoelectronic devices from armchair graphene nanoribbons, which  possess a large  bandgap opening at  the  $\Gamma$-point~\cite{han-prl07,zhou-nature07,xia-nanoletters10,chen-nature15,ugeda-nature14,brey06}.
In these armchair graphene nanoribbon devices, bandgaps can be created in a desired fashion by precise control of the width of the nanoribbon.

Strain engineering in graphene is promising for straintronic and spintronic  applications. Possible ways of creating strain in graphene is to modify the in-plane and out-of-plane deformations among the carbon atoms. For examples, in plane and out-of-plane deformations in graphene can lead to the formation of relaxed shape graphene~\cite{prabhakar16,prabhakar14,shenoy08}. When graphene is relaxed due to precise edge engineering, the localized eigenstates, created  due to strain engineering, can lead to several observed interesting phenomena such as conduction and valence band crossings, spin hot spots as well as measurements of decoherence time\cite{prabhakar14,prabhakar16,prabhakar17epj,prabhakar15jpcc}.
Large in-plane and out-of-plane deformations can also be made by growing graphene on a flexible substrate, where not only the lattice mismatch between graphene and the substrate induces strain but also the flexible substrate can be bended for the purpose of inducing strain~\cite{ni08,bastos16}.
Ripples and wrinkles in graphene are also of potential interest for  band engineering applications~\cite{lim15,bronsgeest15,cerda03,ryan17,prabhakar14,prabhakar16}.
Edge fictionalization of graphene nanoribbons  by -H, -H$_2$, -O, -Br  can also lead to in-plane and out-of-plane deformations and can be used for engineering of graphene bandgaps~\cite{deepika15,lim15,wagner13}.

%%%%%%%%%%%%%zhao08 Nanoletter
The mechanical properties, e.g. the Poisson ratio, Young's and shear modulii, of graphene have been characterized by using both experimental and theoretical techniques\cite{herrero18,yllanes17,faccio09,jiang09,zhao09,bizao17,shenoy08,sgouros18,hossain18,nicholl17,bowick17,wan17}.
The experimental value of Young's modulus of bulk graphite is $0.02$ TPa~\cite{blakslee70} but its value for stack of graphene sheets is $0.5$ TPa~\cite{frank07}, for a  monolayer of graphene oxide is 0.15 TPa~\cite{gomez08} and  for a free-standing monolayer graphene membrane is  0.1 TPa~\cite{lee08}. In Ref.~\cite{lee08}, authors also reported that the intrinsic breaking strength
of graphene as $10$ GPa.  Ab initio
calculations of graphene find the value of Young's modulus
to be 1.11 TPa in Ref.~\cite{van00} and to be 1.24  TPa in Ref.~\cite{konstantinova06}, where thickness of graphene is 0.34nm. By using the semiempirical nonorthogonal tight-binding (TB) method,  the Young's modulus of graphene is reported as  1.206 TPa~\cite{hernandez98}. In summary, the  values of Young's modulus reported recently are in close agreement between experimental and theoretical studies within numerical errors~\cite{papageorgiou17,falin17,zhu17,wang17,choudhary18}. Hence, it is confirmed that the values of Young's modulus are extremely large in comparison to other conventional semiconductor materials. Therefore, finding the value of graphene Young's modulus in several different conditions is important for the applications in designing the devices for  straintronic, spintronic and optoelectronic  applications.
%%%%%%%%%%%%%%%%%

In this paper, we consider armchair graphene nanoribbons functioning their armchair edges by -H and -H$_2$. We then apply compressive strain to the system along y-direction and tensile strain along x-direction. This leads with possible application connected with opening and closing the  bandgaps. In other words, the armchair graphene nanoribbons can be used for dual purposes of semiconductor and metallic. Within density functional theory calculations, our study show that the graphene and graphene with -H functional group can not induce significant values of out-of-plane deformations while graphene with -H$_2$ functional group does induce large out-of-plane deformations exhibiting the properties of relaxed shape graphene~\cite{shenoy08}. Such strain engineering can lead to tuning the bandgaps from semiconductor to metal and then again switching back to semiconductor. The binding energy calculations of hydrogen on graphene allow to quantify the effect of  strain on the bandengineering of graphene nanoribbons. Finally, we discuss the effect of strain on the Poisson ratio, Young's and shear modulii of graphene nanoribbons.

The paper is organized as follows. In Sec.~\ref{theoretical-model}, we provide computational details of density functional theory calculations for generating strain on graphene nanoribbon.  In Sec.~\ref{rd}, we present the results for bandengineering, poisson ratio, Young's and shear modulii of graphene, graphene edge functioning with H and H$_2$ .  Finally, in Sec.~\ref{conclusion}, we summarize our results.

\begin{figure}
\includegraphics[width=8.5cm,height=3cm]{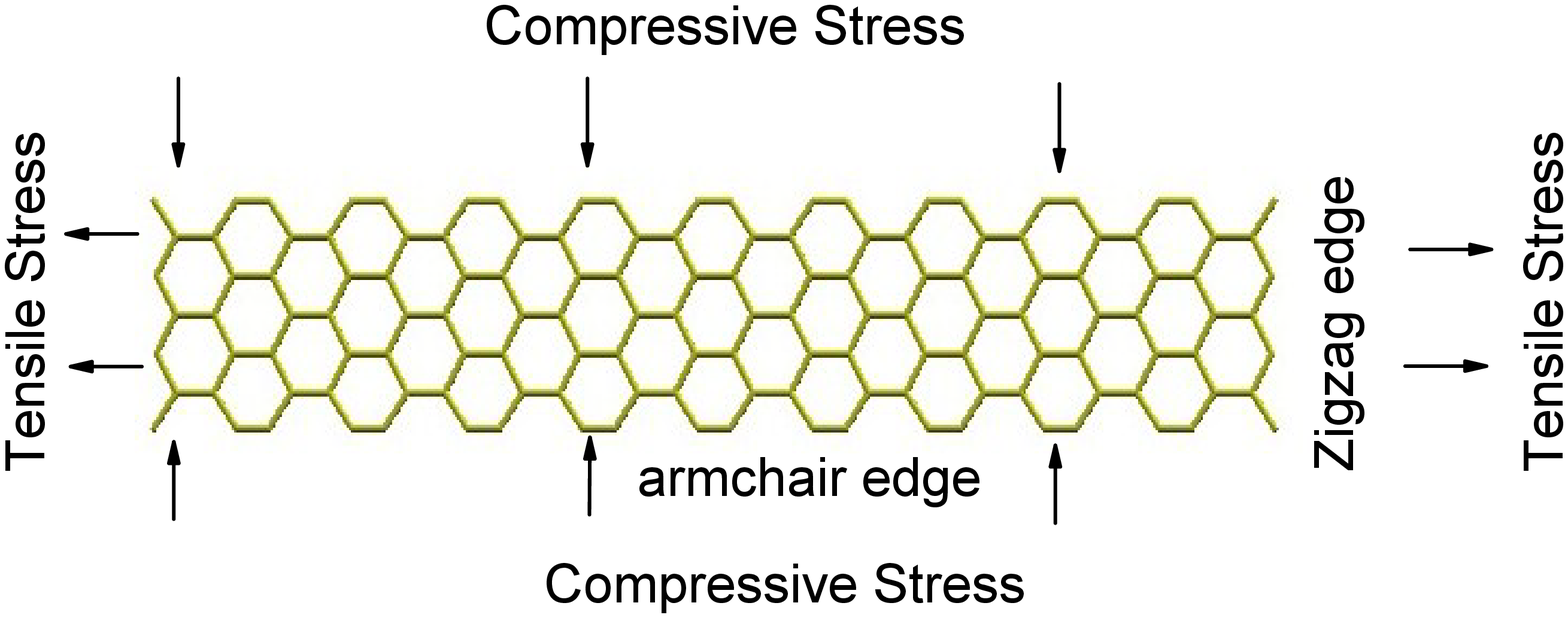}
\caption{\label{fig1} (color online) Schematics of 7 zigzag edges of armchair graphene nanoribbon.  We apply compressive stress at the armchair edge while tensile stress at the zigzag edge. Such type of strain engineering allow us to investigate the influence of strain on the  bandengineering of graphene nanoribbon.
}
\end{figure}

%%%%%%%%%%%%%%%%%%%%%%%%%%%%%%%%%%%
%\begin{figure*}
%\includegraphics[width=18cm,height=7cm]{fig1-7aGNR-SC.eps}
%\caption{\label{fig1-7aGNR-SC}
%}
%\end{figure*}
%%%%%%%%%%%%
\begin{figure}
\includegraphics[width=8.5cm,height=4cm]{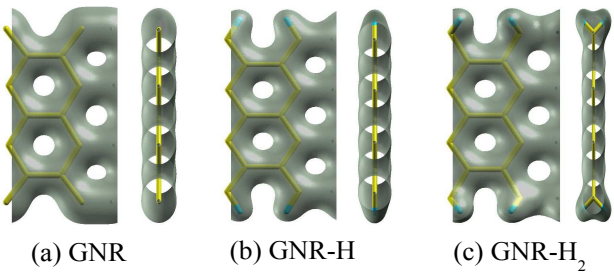}
\caption{\label{fig2} (color online) Isosurface charge distributions of armchair graphene nanoribbon for (a) bare graphene (b) armchair edge functioning with hydrogen ($H$) and (c) armchair edge functioning with hydrogen molecule ($-H_2$).
}
\end{figure}
%%%%%%%%%%%%
%%%%%%%%%%%%
\begin{figure*}
\includegraphics[width=18cm,height=8cm]{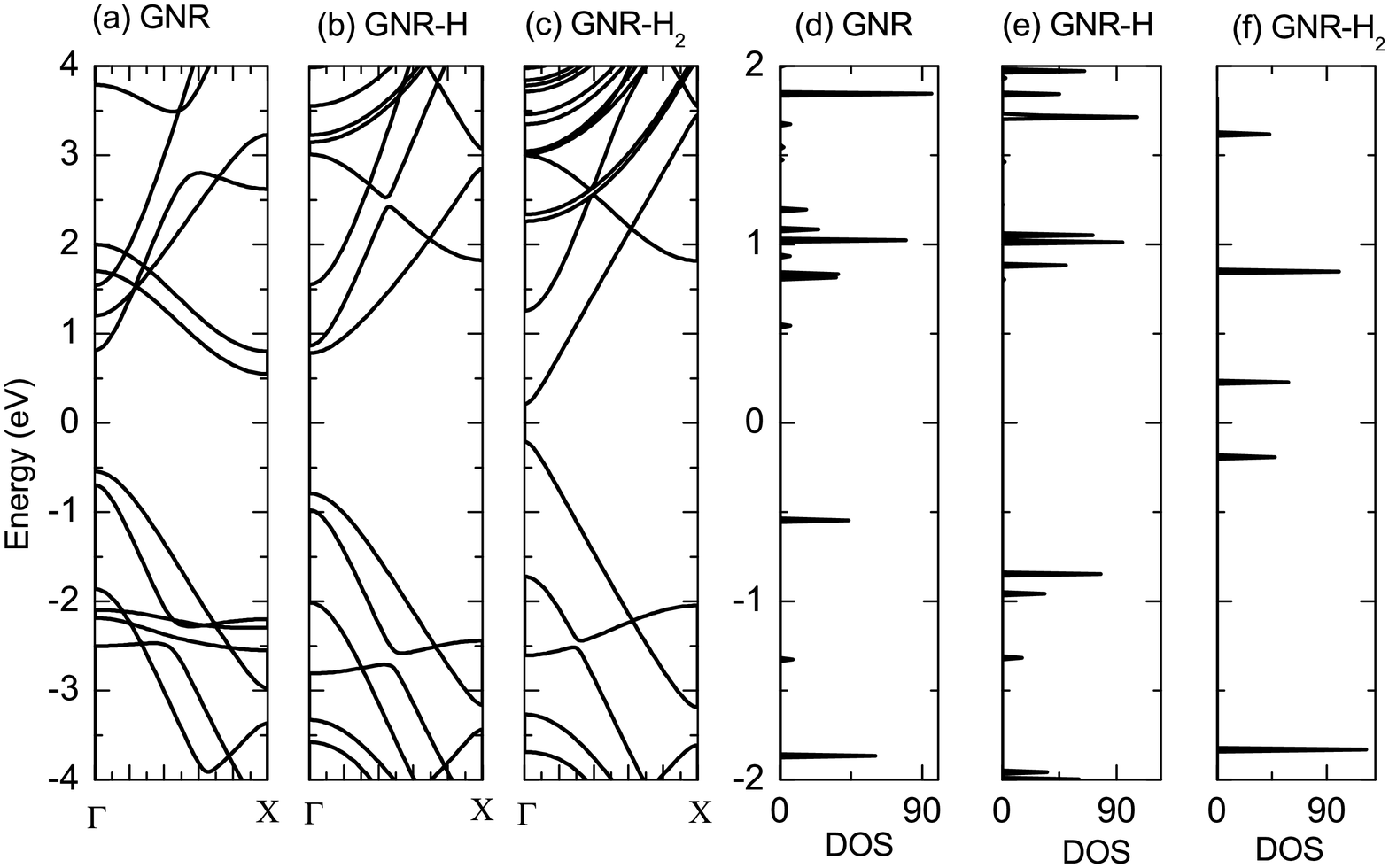}
\caption{\label{fig3banddos} (color online) Band structures and Density of States (DOS) of graphene nanoribbon, graphene nanoribbon armchair edge functioning with H, and graphene nanoribbon armchair edge functioning with $H_2$.
}
\end{figure*}
%%%%%%%%%%%%
\begin{figure}
\includegraphics[width=8.5cm,height=6cm]{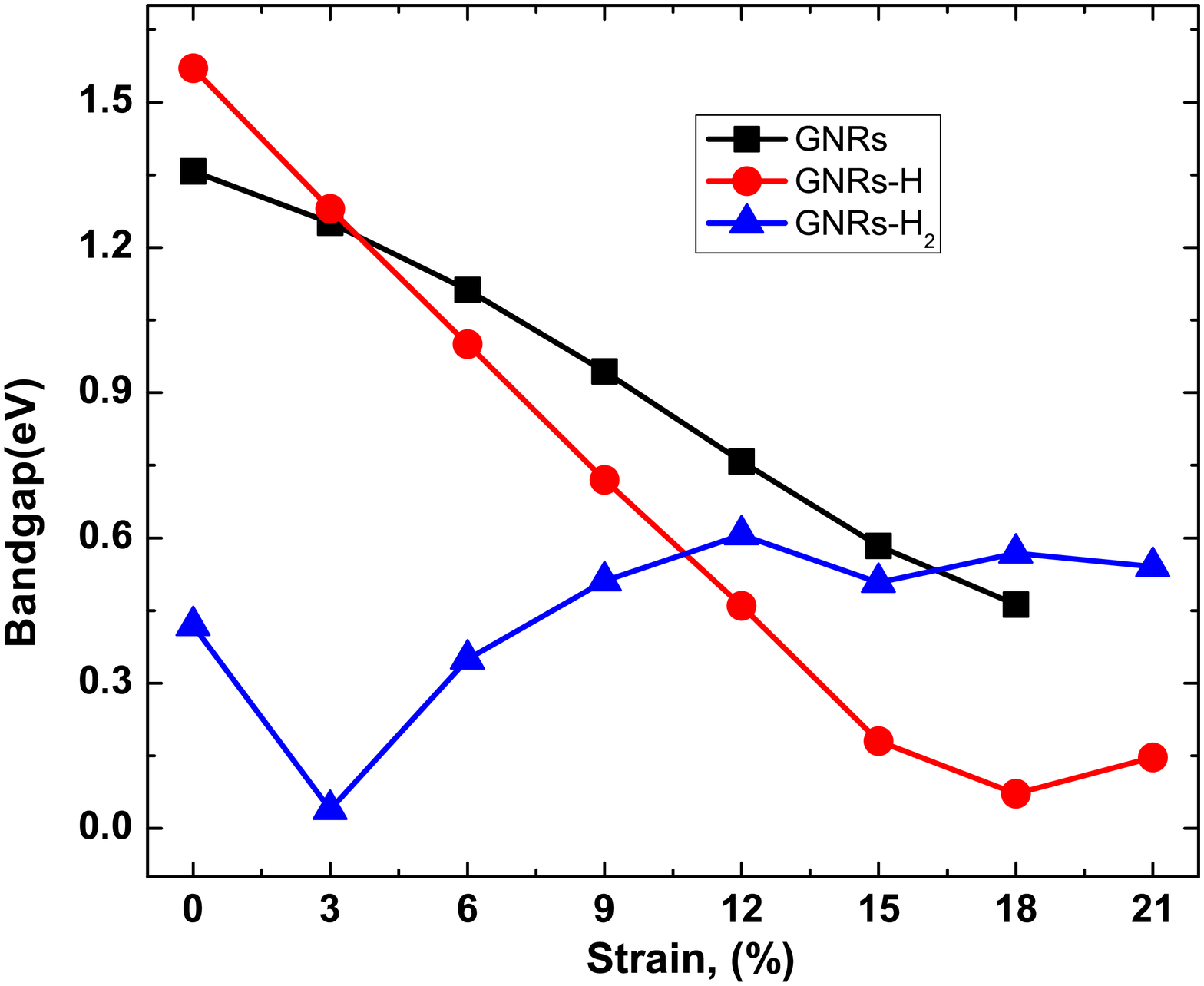}
\caption{\label{fig4bandEng} (color online) Band engineering in the armchair graphene nanoribbon with the application of strain in three cases: (a) graphene nanoribbon, (b) graphene nanoribbon functioning with hydrogen at the armchair edge and (c) graphene nanoribbon functioning with $H_2$ molecule at the armchair edge. We see that the bandgap completely closes at the particular value of strain for the case of graphene nanoribbons functioning with H and H$_2$.
}
\end{figure}
\begin{figure}
\includegraphics[width=7cm,height=7cm]{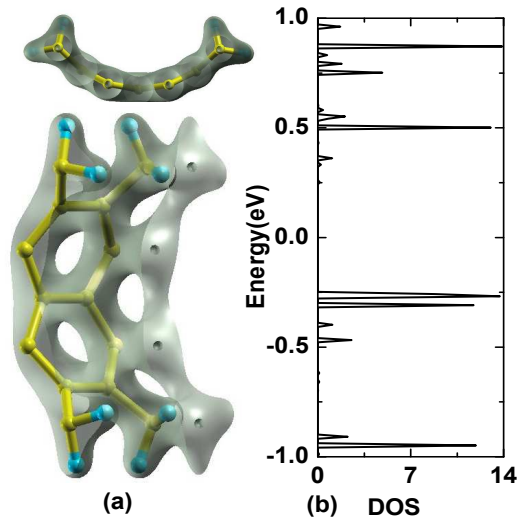}
\caption{\label{fig-relaxed-H2-15} (color online) Isosurface charge distribution in (a) and energy vs density of states in (b)  of armchair graphene nanoribbon for  armchair edge functioning with hydrogen molecule ($-H_2$) under $15\%$ compressive strain. Notice that the out-of-plane deformations are significantly large that provided the relaxed shape graphene nanoribbon similar to Ref.~\cite{shenoy08,prabhakar14}. 
}
\end{figure}

\begin{figure}
\includegraphics[width=8.5cm,height=4cm]{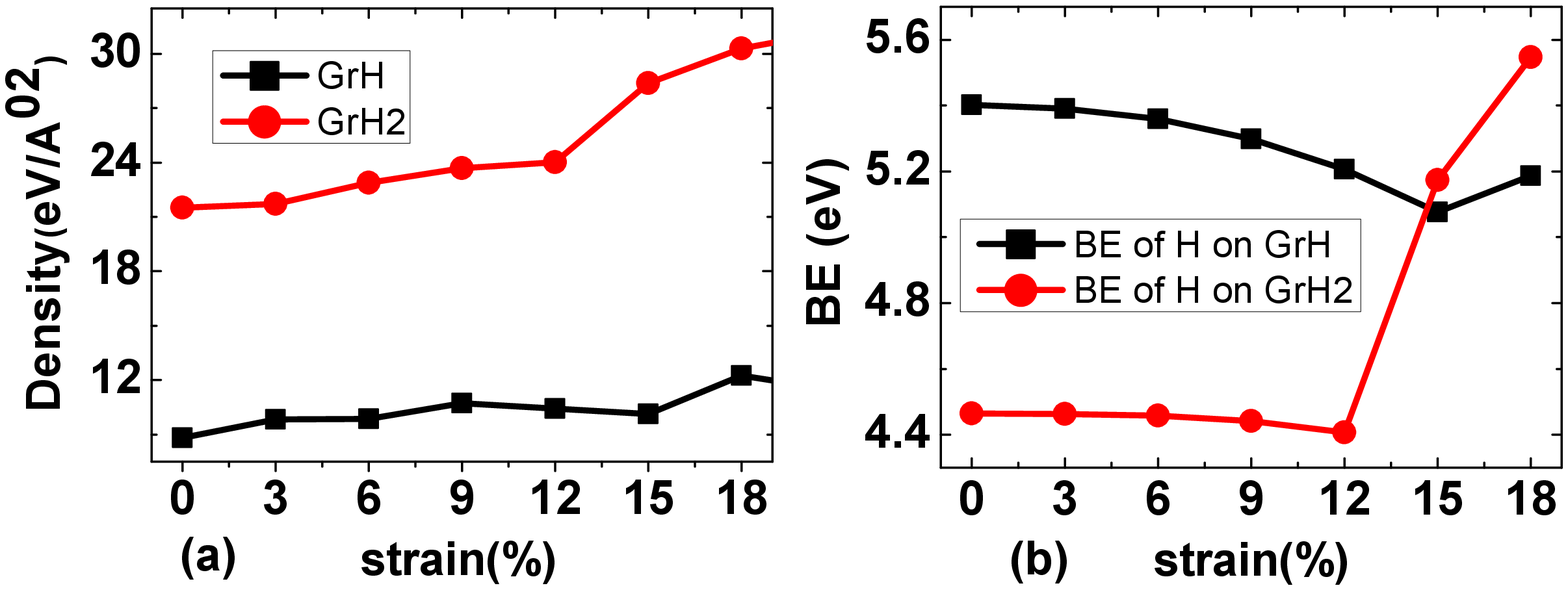}
\caption{\label{fig6BE} (color online) The variation of relative total energy density vs strain in (a) and binding energy of hydrogen on armchair edge of graphene vs strain in (b). The relative total energy is obtained by subtracting the total energy of the graphene nanoribbon from the total energy of graphene nanoribbon with $-H$ (diamond-black in Fig.~\ref{fig6BE}(a)) and $-H_2$ (circle-red in Fig.~\ref{fig6BE}(a)).
}
\end{figure}

\begin{figure}
\includegraphics[width=8.5cm,height=4cm]{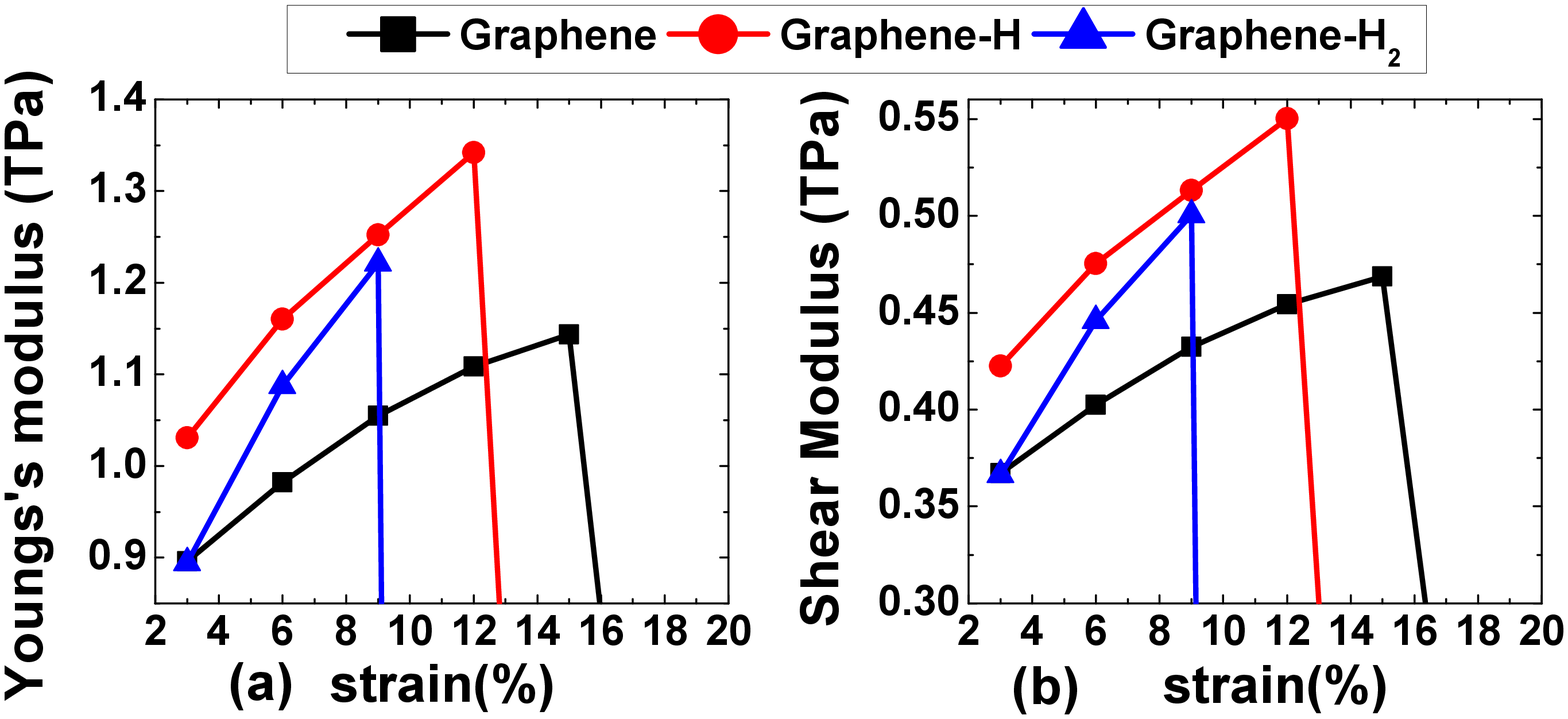}
\caption{\label{fig7} (color online) The variation of Young's modulus with respect to strain in (a) and shear modulus vs strain in (b). Evidently, Young's and shear modulii increase as we increase the strain until they reach their saturation values. The Young's modulus is calculated by using Eq.~\ref{Y} and shear modulus is calculated by the use of Eq.~\ref{G}. Here we consider the Poisson ratio, $\nu = 0.22$.
}
\end{figure}

\section{Computational Methods}\label{theoretical-model}
Density Functional Theory (DFT)  calculations for 7 zigzag edges of armchair graphene nanoribbons are performed  in the Quantum Espresso software package~\cite{QE09}, where periodic boundary conditions are implemented in the simulations.  Ultrasoft pseudopotentials and plane wave basis set with a kinetic energy and charge density  cut-off at 100Ry and 800Ry are used. We include exchange and correlation effects within the Perdew-Burke-Ernzerhof (PBE) Functional~\cite{perdew96}.
Van der Waals interactions are also included with the Semiempirical Grimme's DFT-D2 corrections term~\cite{grimme06}.
We use orthorhombic  4.34${\AA}$ $\times$ 24.78${\AA}$ $\times$ 15.01${\AA}$ size of supercell that contains 7 zigzag edges of an armchair graphene nanoribbon. In the supercell, we have 14 atoms for the graphene nanoribbon, 18 atoms for the graphene nanoribbon with armchair edge functioning with H and  22 atoms for the graphene nanoribbon with armchair edge functioning with H$_2$.  During geometry optimization, all atoms in all the three directions and x-axis of the supercell are fully relaxed until the forces on atoms are smaller than 0.01 eV/$\AA$. We have tested several k-point samplings and calculations are performed at (6,1,3) $k$-points sampling  fulfilling the above convergence criteria. The optimized lattice parameters along the x-direction are 4.34${\AA}$ for graphene, 4.29${\AA}$ for graphene with armchair edge functioning with H, 4.30${\AA}$ for graphene with armchair edge functioning with H$_2$. The optimized widths (distance between C-C atoms) of 7 zigzag edges of armchair graphene nanoribbons  are 7.1353 ${\AA}$ (bare graphene), 7.33${\AA}$(graphene edge functioning with H), 7.56 ${\AA}$ (graphene edge functioning with H$_2$).   The XCRYSDEN program was used to draw the molecular structure~\cite{xcrysden}. For strain engineering of graphene nanoribbons, we have applied the compressive stress through the armchair edge and then frozen the y-coordinates of the armchair edge atoms while  allowing the other coordinates to be fully relaxed. We also fully relaxed the x-axis of the supercell.

\section{Results and Discussions}\label{rd}
Bandgaps of the armchair graphene nanoribbon depend on the width of the ribbon. If the armchair nanoribbon contains an odd number of zigzag edges then the armchair graphene nanoribbon possesses a finite bandgap opening at the $\Gamma$ point and interesting  for making  semiconductor devices. For an even number of zigzag edges, the armchair graphene nanoribbon has a zero band gap opening and thus possesses metallic properties. When strain is implemented in the armchair graphene nanoribbon with  odd number of zigzag edges, then such a nanoribbon possesses the properties of both semiconductor and metal.   In this paper, we are interested in tuning the bandgap from a semiconducting to metallic and then bringing it back to the semiconductor with the application of strain engineering. Hence, we consider the 7 zigzag edges of the armchair graphene nanoribbon and then apply the compressive stress at the armchair edge. The armchair ribbon is also allowed to elongate in the x-direction. The schematics of the strained armchair graphene nanoribbon is shown in Fig.~\ref{fig1}.

In Fig.~\ref{fig2}, we have plotted the isosurface charge distributions on 7 zigzag edges of armchair GNRs for three different cases: (i) bare GNRs (Fig.~\ref{fig2} (a)), (ii) GNRs with H at the armchair edge (Fig.~\ref{fig2} (b)) and (iii) GNRs with H2 at the armchair edge. As can be seen in Fig.~\ref{fig2}, the distributions of isosurface charge in these GNRs are different because different functional groups in GNRs induce slightly different strains. Hence bandgap opening of these ribbons are expected to be different. The bandstructures and density of states of these GNRs are shown in Fig.~\ref{fig3banddos}. At $\Gamma$ point, we find that the bandgaps of  GNRs are 1.36eV for bare GNRs, 1.57eV for GNRs with H and 0.42eV for GNRs with H$_2$. Since all the GNRs has 7 zigzag edges but different bandgap openings, it clearly indicates that the attachment of hydrogen at the armchair edge plays an important role in the bandgap engineering. For example, the attachment of hydrogen changes the width, (d(C-C)), of the ribbon (7.13${\AA}$ vs 7.32${\AA}$ vs 7.54${\AA}$) and also induces the in-plane and out-of-plane deformations~\cite{wagner13}.

The tuning of the bandgaps from semiconductor to metal and bring it back to semiconductor can be achieved by applying external strain through the armchair edge, as shown schematically in Fig.\ref{fig1}. The bandgap engineering with respect to strain is shown in Fig.~\ref{fig4bandEng}. In Fig.~\ref{fig4bandEng} (diamond), we find that graphene always stays as a semiconductor even for large values of strain. On the other hand,  in Fig.~\ref{fig4bandEng}(circle) the armchair edge passivated with hydrogen become metallic (i.e., no bandgap) approximately at $18\%$ of applied strain. The bandgap closing can be achieved even at much smaller values of strain (approximately at $3\%$ of strain) for the case of graphene nanoribbon passivated with H$_2$. The bandgap closing can be seen due to the effect of strain because applied strain decreases the width of the ribbon. When the ribbon width becomes closer to the width of the 6 (even number) zigzag edge armchair graphene nanoribbon then the metallic behavior in the band engineering of graphene nanoribbon can be seen~\cite{rizzo18,groning18}.

Using finite element method simulations, authors in Ref.~\cite{shenoy08,prabhakar14} have shown that the relaxed shape graphene nanoribbon has potential interest for graphene-based strain engineering devices~\cite{shenoy08} as well as for the observation of  quantum phenomena e.g. discrete energy levels, pseudospin decoherence time and nontrivial topological phases~\cite{prabhakar16,groning18,rizzo18,cao17}. When the out-of-plane deformations are large as can be seen in Fig.~\ref{fig-relaxed-H2-15}(a) for the case of graphene passivated with hydrogen, we observe a relaxed shape graphene nanoribbon. In Fig.~\ref{fig-relaxed-H2-15}(b) we have plotted the density of states and find the bandgap reopening in the graphene nanoribbon.

Now we turn to the discussions of mechanical properties observed in the armchair graphene nanoribbon due to externally applied compressive strain from the y-direction and tensile strain in the x-direction.

In Fig.~\ref{fig6BE}(a), we plot the relative total energy density  (i.e., energy density difference relative to  graphene). This total energy is used to calculate the Young's modulus,  shear modulus and the binding energy of hydrogen. In this figure, we find that the total energy density increases significantly for the case of graphene passivated with $-H_2$ molecules (circles) at $12\%$ of strain which provides an indication of relaxed shape graphene~\cite{shenoy08}. In Fig.~\ref{fig6BE} (b), we plot the binding energy of hydrogen in the graphene which allows to estimate the strain generated by hydrogen on the graphene nanoribbon. The binding energy is found by using the expression:~\cite{prabhakar17sr}
\begin{equation}
BE = U[Gr(H, H_2)] - U[Gr] - U[H],
\end{equation}
where BE is the binding energy, $U[Gr(H, H_2)]$ is the total energy of graphene with passivation layer, $U[Gr]$ is the total energy of graphene and $U[H]$ is the total energy of hydrogen in vacuum.   As can be seen in Fig.~\ref{fig6BE} (b), the binding energy of hydrogen in the graphene passivated with $-H_2$ enhances significantly after $12\%$ of strain, which provides a clear indication of relaxed shape graphene (i.e., very large out-of-plane deformations).

Finally, we investigate the effect of strain on the Young's modulus, Poisson ratio and shear modulus of graphene nnaoribbon. The Young's modulus is calculated by using the expression~\cite{faccio09}
\begin{equation}
Y = \frac{1}{A_0}\frac{d^2U}{d\varepsilon_{y}^2},\label{Y}
\end{equation}
where $A_0$ is the optimized area without imposing any strain, $U$ is the total energy, and $\varepsilon_{y}$ is the externally applied compressive strain. In Fig.~\ref{fig7}(a), we find that the Young's modulus of graphene passivated with $H_2$ is larger than the Young's modulus of graphene and graphene passivated with H. Young's modulus also increases as we increase the strain and then sharply decreases providing an indication of saturation value of the Young's modulus. The numerical values of Young's modulus presented in  Fig.~\ref{fig7}(a) are consistent with the experimental and theoretical values reported in the literature.~\cite{faccio09,jiang09,zhao09,bizao17,shenoy08,sgouros18,hossain18,nicholl17,bowick17}

The Poisson ratio is another important physical quantity characterizing the mechanical properties of materials~\cite{burmistrov18,jiang09,faccio09,zhao09}.
By applying compressive external strain, $\varepsilon_y$, in the graphene in the  y-direction, and by using the supercell of DFT calculations to record the resulted strain in the x-direction, $\varepsilon_x$, we find the Poisson ratio from the expression, $\nu=\varepsilon_x/\varepsilon_y$ as $\nu=0.25$ for graphene, $\nu=0.25$ for graphene passivated with H and $\nu=0.22$ for graphene passivated with H$_2$. Since the Poisson ratios for all three cases are almost the same ($\nu = 0.25$ (Gr), $0.25$ (Gr-H), and  $0.22$ (Gr-H$_2$)), we can draw a conclusion that the Poisson ratio of graphene is  unaffected by the passivation of -H and H$_2$ at the armchair edge.

After calculating the Young's modulus and Poisson ratio, we can finally find the shear modulus, $G$, by using the expression,
\begin{equation}
G = \frac{Y}{2(1+\nu)}.\label{G}
\end{equation}
The shear modulus with respect to applied compressive strain is plotted in Fig.~\ref{fig7}(b). Similar to Young's modulus, shear modulus also increases as we increase the strain until it reaches to the saturation value.

\section{Conclusion}\label{conclusion}
In summary, we have studied the effect of strain on the band engineering (e.g., semiconductor vs mettalic) and the mechanical properties (e.g. Young's modulus, shear modulus and Poisson ratio) of armchair graphene nanoribbons passivated with hydrogen and hydrogen molecules at the armchair edges.  We have shown that the strain can be used  for tuning the bandgap from semiconductor to metallic,  and then strain can again be used to tune the bandgaps from metallic characteristics to semiconductor. We have also shown that at the large values of strain in graphene nanoribbons passivated with hydrogen molecules, large out-of-plane deformations in graphene nanoribbon can show the properties of relaxed shape graphene. We have quantified the strain induced by hydrogen molecules on graphene nanoribbons in terms of the binding energy. Finally, we have shown that the Young's and shear modulii increase as we increase the strain until they reach to their maximum values while the Poisson ratio is independent to the passivation materials. Our findings of bandgap tuning and exploration of mechanical properties can be used in designing novel optoelectronic devices, as well as for straintronic applications.

\begin{acknowledgments}
The authors acknowledge the support made by Canada Research Chair Program and National Research Council of Canada.  The computations were performed utilizing the facilities available on the SHARCNET Supercomputer(www.shracnet.ca) and Compute Canada supercomputer (www.computecanada.ca).
\end{acknowledgments}

%\bibliography{bib18}
%Merlin.mbs v4.21 2009-07-09.
%

\end{document}